# Obstacle evasion in free-space optical communications utilizing Airy beams

Guoxuan Zhu[1,†], Yuanhui Wen[1,†], Xiong Wu[1,†], Yujie Chen[1,*], Jie Liu[1], and Siyuan Yu[1,2]

[1]State Key Laboratory of Optoelectronic Materials and Technologies, School of Electronics and Information Technology, Sun Yat-sen University, Guangzhou 510275, China

[2]Photonics Group, Merchant Venturers School of Engineering, University of Bristol, Bristol BS8 1UB, UK

[†]These authors contributed equally to this work

*Corresponding author: chenyj69@mail.sysu.edu.cn

**A high speed free-space optical communication system capable of self-bending signal transmission around line-of-sight obstacles is proposed and demonstrated. Airy beams are generated and controlled to achieve different propagating trajectories, and the signal transmission characteristics of these beams around the obstacle are investigated. Our results confirm that, by optimising their ballistic trajectories, Airy beams are able to bypass obstacles with more signal energy and thus improve the communication performance compared with normal Gaussian beams.**

Free space light fields propagating along curved trajectories, also known as accelerating beams, have been intensively investigated in recent years since the first demonstration of an Airy beam within the context of optics in 2007 [1]. More and more kinds of accelerating beams are subsequently proposed and generated moving along various trajectories including convex [2] and nonconvex ones [3] in both two-dimensional and three-dimensional space [4-6]. Nonetheless, Airy beams, as the simplest kind of accelerating beams with peculiar properties including not only self-bending but also self-healing and non-diffracting [7,8], are attractive for a variety of potential applications, such as micromanipulation [9,10], filamentation [11], vacuum electron acceleration [12], light bullets [13,14], micromachining [15], plasmonics [16,17], optical routing [18] and imaging [19,20]. In addition, Airy beams as a new type of structured light beams in addition to Gaussian beams, Bessel beams, vortex beams, etc. are also introduced to the field of free-space optical communication, mainly due to their self-healing property, which could help the main lobe of these beams resist small obstacles or atmosphere turbulences [21,22]. Transmission of spatial image signal encoded in the Fourier plane has been presented through a 4f telescope system [23]. However, truly high speed free-space communication based on Airy beams is not yet demonstrated and whether the peculiar ballistic dynamics of Airy beams is beneficial for evading obstacle and thus improving the communication performance still needs to be assessed.

In this Letter, we proposed and demonstrated an Airy-based high speed optical communication system as well as evaluating its communication performance around obstacles. One-dimensional truncated Airy beams are employed in this case for evading transverse obstacles in both simulation and experiment. A comparison of Airy beams and normal Gaussian beams in terms of their obstacle evading performance has been carried out, which shows that Airy beams with optimized ballistic trajectories can avoid obstacles almost the size of the initial beam waist radius while the normal Gaussian beam is mostly blocked under the same condition. High speed optical communication is further implemented and the communication performance is also evaluated to confirm the improvement afforded by Airy beams over Gaussian beams. Our finding suggests that Airy beams as a new type of structured light beams might be useful in high speed free-space communication, especially where the beams need to overcome small line-of-sight obstacles within the communication link.

The established communication system based on Airy beams is shown in Fig. 1. A 28-Gbaud QPSK signal is modulated onto the 1550nm wavelength light beam by an arbitrary waveform generator (AWG) and amplified by an Er-doped fiber amplifier (EDFA). This light beam is then collimated to a beam waist radius (denoted as $\omega_0$) about 2mm in free space and subsequently modulated by a phase-only spatial light modulator (SLM1) for Airy beam generation. The one-dimensional Airy beam can be generated by loading a cubic phase pattern, $\alpha \cdot (x-x_0)^3$, on the SLM1, where $\alpha$ is the cubic parameter, $x$ is the transverse coordinate and $x_0$ is the transverse shifting parameter. The total insertion loss of this reflective SLM (PLUTO-2-TELCO-013, HOLOEYE) is around 0.6dB including the diffraction and response loss. After performing Fourier transformation with a lens, the Airy beam is generated near the back focal plane of the lens with a focal length of 20cm. Apart from generating Airy beams, the SLM1 is also capable of adjusting the ballistic trajectory of Airy beams so as to get over a preset obstacle with more bypassing energy. Two approaches are employed here, including (a) changing the cubic parameter $\alpha$ and (b) shifting the phase pattern by adjusting $x_0$. Approach (a) can optimize the degree of bending of the Airy beam and approach (b) can optimize the emission angle of the ballistic trajectory. Both approaches can be realized by simply loading patterns on SLM without manually adjusting other optical devices.

An opaque thin plate near the focal plane of the lens is used as the obstacle to transversely block the signal beam. The propagation trajectory of the signal beam is monitored by a camera placed on a slider after a 4f imaging system. The focal length of the lenses in this imaging 4$f$ system is 25cm. Since the performance of an optical receiver in capturing optical power is related to their receiving aperture, a pinhole is put right after the receiving lens to mimic a finite receiving aperture in practice. The pinhole has the same aperture as the transmitter ($\omega_0$) and has been aligned with a Gaussian beam. The received energy is detected by a power meter. At the same time, the Airy mode is demodulated by the SLM2 and the signals are collected by a collimator and a single-mode-fiber pigtail. For directly evaluating the obstacle evasion ability, the PD is put parallel to the SLM2 to avoid the influences caused by SLM2 mode conversion and misalignment issues. The signals then go through a QPSK receiving system and are processed offline. An amplified spontaneous emission source and a variable optical attenuator are applied to act as noise with around 0.1nm bandwidth and -40dBm power at 1550nm, while the received signal light power without obstacle is around -5dBm. The signal processing results reflect the performance of the entire communication scheme.

In order to intuitively illustrate the impacts of the two adjusting parameters of the cubic phase mentioned before on the beam trajectory, the propagation of a series of Airy beams with different combination of the two parameters has been simulated by the beam propagation method (BPM). The optical field of an objective plane at $z=z_0$ can be calculated based on the given initial optical field and the corresponding propagating distance $z_0$. After sweeping the propagating distance $z_0$ and record the field distribution on each objective plane, the two-dimensional (2D) trajectory profile can be acquired.

Figure 2(a) presents the 2D simulated trajectories of the Airy beam with $α$ and $x_0$ shown in each sub-panel respectively, where $x$ is the transverse direction, $z$ is the beam propagation direction, and $z=0$ is the focal plane. The idea of the Airy beam communication scheme is utilizing these highly-tailorable propagating trajectories to overcome irregular-shaped obstacles. Since the beam propagation method is based on paraxial approximation, the simulated trajectories range is from $z=-10$cm to $z=20$cm. Focusing on the region between $z=-10$cm and $z=10$cm, Fig. 2(a1) and 2(a3) indicate that the trajectories of Airy beams modulated by the same $α$ and the opposite $x_0$ are symmetric about the focal plane while Fig. 2(a1) and 2(a4) indicate that the trajectories modulated by the opposite $α$ and the same $x_0$ are centrally symmetric about the focal point. As a result, the trajectories modulated by the opposite $α$ and $x_0$ are symmetric about the plane $x=0$. Considering these symmetric properties of Airy beams' trajectories, the sign of $α$ and $x_0$ can be selected according to the location of the obstacle [e.g. when obstacle is located on the lower-left side, the signs of $α$ and $x_0$ should be minus as shown in Fig. 2(a6)].

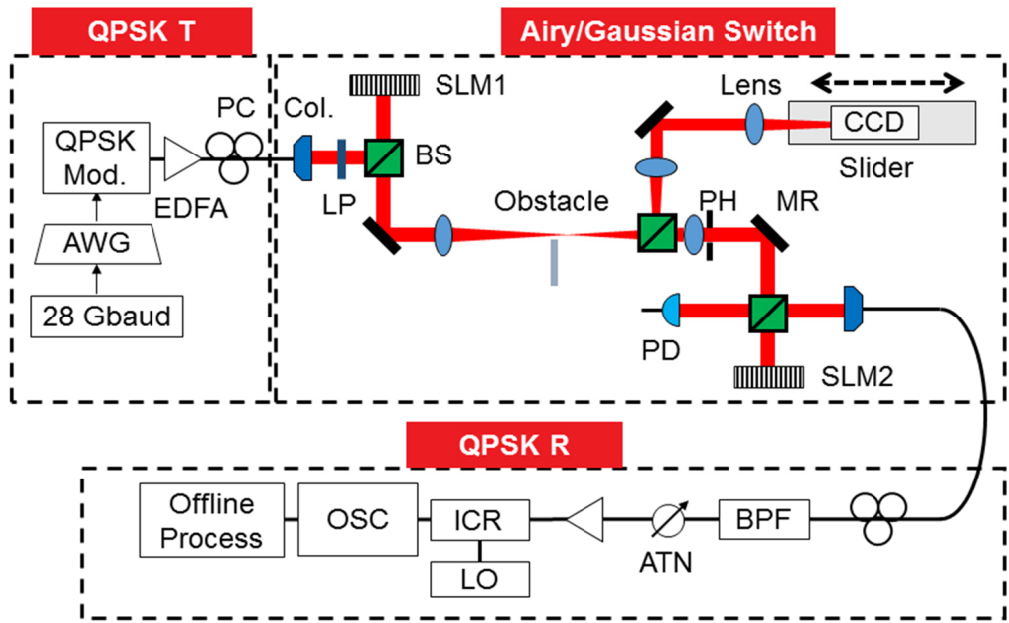

Fig. 1. Experiment setup of the communication system overcomes obstacles with Airy and Gaussian beams. AWG: arbitrary waveform generator; QPSK: quadrature phase-shift keying; EDFA: Erbium-doped fiber amplifier; PC: polarization controller; Col.: collimator; LP: linear polarizer; SLM: spatial light modulator; BS: beam splitter; PH: pinhole; MR: mirror; PM: power meter; BPF: band-pass filter; ASE: amplified spontaneous emission; VOA: variable optical attenuator; ICR: integrated coherent receiver; LO: local oscillator; OSC: oscilloscope.

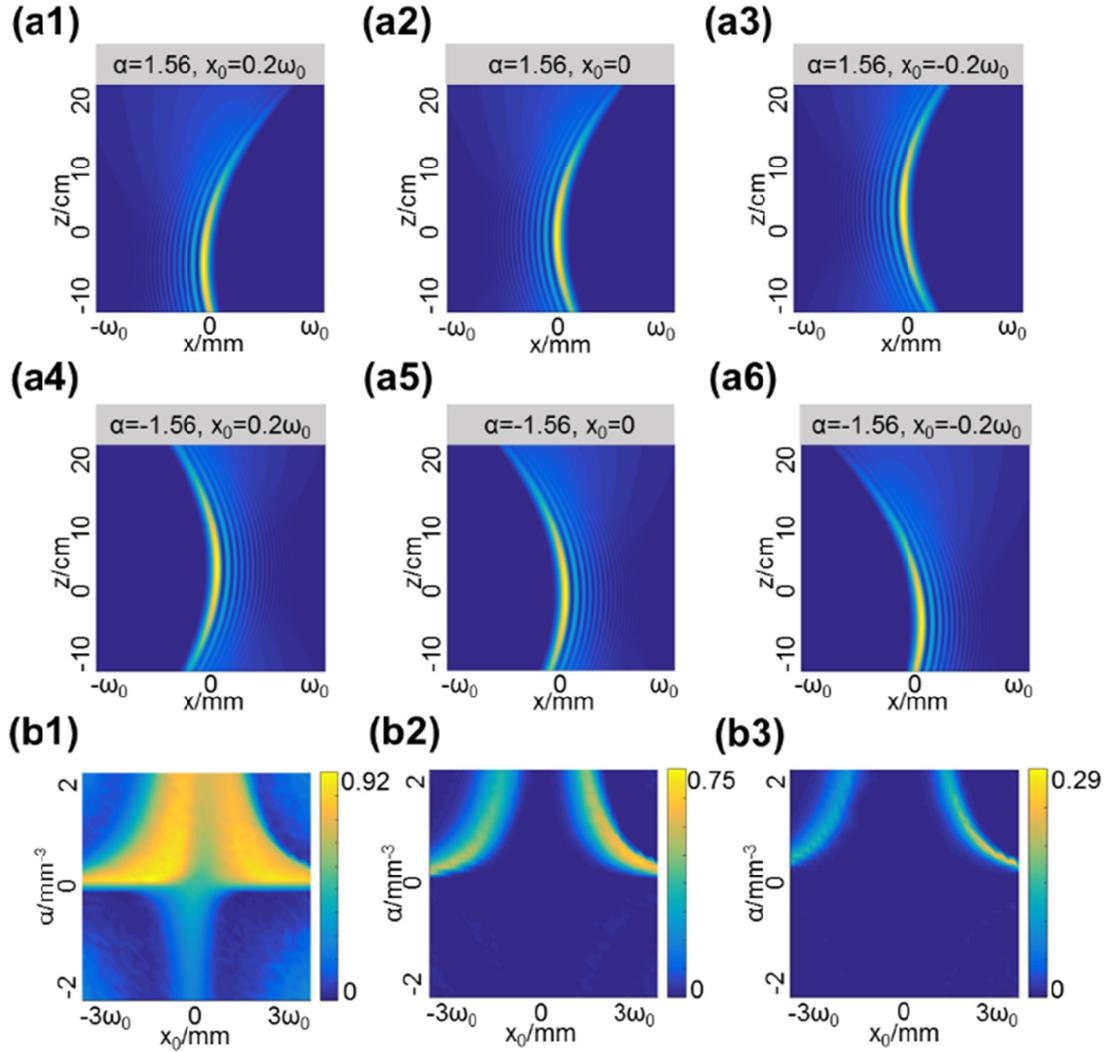

Fig. 2. (a) Simulation results of the ballistic dynamics of Airy beams with different parameters $\alpha$ and $x_0$, the unit of $\alpha$ is $mm^{-3}$. (b) Experimental results of the bypassing power of the Airy beams with different parameters $\alpha$ and $x_0$ while an obstacle on the focal plane blocks the light beam from $x=-\omega_0$ to (b1) $x=0$, (b2) $x=\omega_0/2$ and (b3) $x=\omega_0$.

As shown in Fig. 2(a), adjusting $\alpha$ and $x_0$ are beneficial for dispersing the beam power to specific area to overcome the obstacle. However, restricted by the finite aperture (pinhole) at the receiver, the values of these two modulation parameters should be carefully selected to avoid dispersing power outside this receiving aperture located at the plane $z=20cm$. The experimental results of the received optical power (normalized by the incident power) related to the two modulation parameters are shown in Fig. 2(b). The obstacle is settled on the focal plane and blocks the beam from $x=-\omega_0$ to $x=0$, $x=\omega_0/2$ and $x=\omega_0$, respectively. It should be noted that the maximum normalized power is around 0.92. This is mainly due to the diffraction loss of SLM1 because the diffraction efficiency of SLM is related to the phase variation of the cubic phase for Airy beam generation, which changes sharply in order to overcome a large obstacle. As shown in Fig. 2(b1), the low receiving power with minus $\alpha$ is caused by the obstacle and the low receiving power with large $\alpha$ and absolute $x_0$ is caused by the pinhole. As the obstacle increasingly blocks the beam along the plane, it is necessary to apply larger $\alpha$ and $x_0$ to evade the obstacle, and finally the power within the received aperture inevitably gets lower. Fig. 2(b3) indicates that the ultimate obstacle size that Airy beams can get around is close to the original beam waist radius and the receiving aperture.

To illustrate the benefit of utilizing the ballistic trajectories of Airy beams for getting around obstacles, we compare the obstacle-evading ability of the Airy beams with normal Gaussian beams, as shown in Fig. 3. The Gaussian beam is generated simply by replacing the previous cubic phase with a quadratic lens phase so that the beam size is adjustable. It should be noted that the normal Gaussian beam here is symmetric about the beam axis as usual without beam steering. Fig. 3(a) shows three typical trajectories of the original Gaussian beam, deformed Gaussian beam by adjusting the lens phase and deformed Airy beam. Each trajectory is processed and obtained from eleven images of beam patterns recorded by the camera on the slider as shown in Fig. 1, and the beam patterns on the focal plane are presented in Fig. 3(b) particularly. The longitudinal interval length between the adjacent recorded images is 10 mm, so these eleven images exhibit the beam trajectory propagating from $z$=-50 mm to $z$=50 mm. With an obstacle put on the focal plane as indicated by the dashed rectangle, the original Gaussian beam in Fig. 3(a1) is mostly blocked. By adding a lens mask to the original Gaussian beam to adjust its beam size, the deformed Gaussian beam can partially get over the obstacle. While for Airy beams with optimized ballistic trajectories, they can shift most of their energy to the upper side of the propagation region and thus mostly circumvent the obstacle. By altering the position of the obstacles, it is found that the Airy beam can overcome larger obstacle with more bypassing energy than the normal Gaussian beams. The relationship between optimal receiving power and the location of obstacle is shown in Fig. 3(c). The obstacle location is marked by the position of the top end of the obstacle while the lower part of the plane is totally blocked. When the transversely blocked region is less than half the plane, both deformed Gaussian and Airy beam can overcome the obstacle. However, when the blocked region is more than half the plane, the power of deformed common Gaussian beam decreases sharply while the Airy beam can still transmit until the obstacle blocks the whole beam waist. The gain in received power by using the Airy beam compared with the normal Gaussian beam is up to 9 dB.

Both Airy and Gaussian beams are examined in a free space communication system with an obstacle put on the focal plane $z$ = 0 and the plane $z$ = 50 mm after the focal plane respectively. The communication performance is evaluated by the error vector magnitude (EVM) between the received signals after DSP and the respective ideal signals [24]. The value of EVM characterizes the dispersion of the signal points on the constellation diagram and a larger value means worse performance. The EVM result is related to the signal to noise ratio (SNR) in principle. Here, we had added -40dBm noise to mimic the practical situation and the SNR without obstacle is around -35dB.

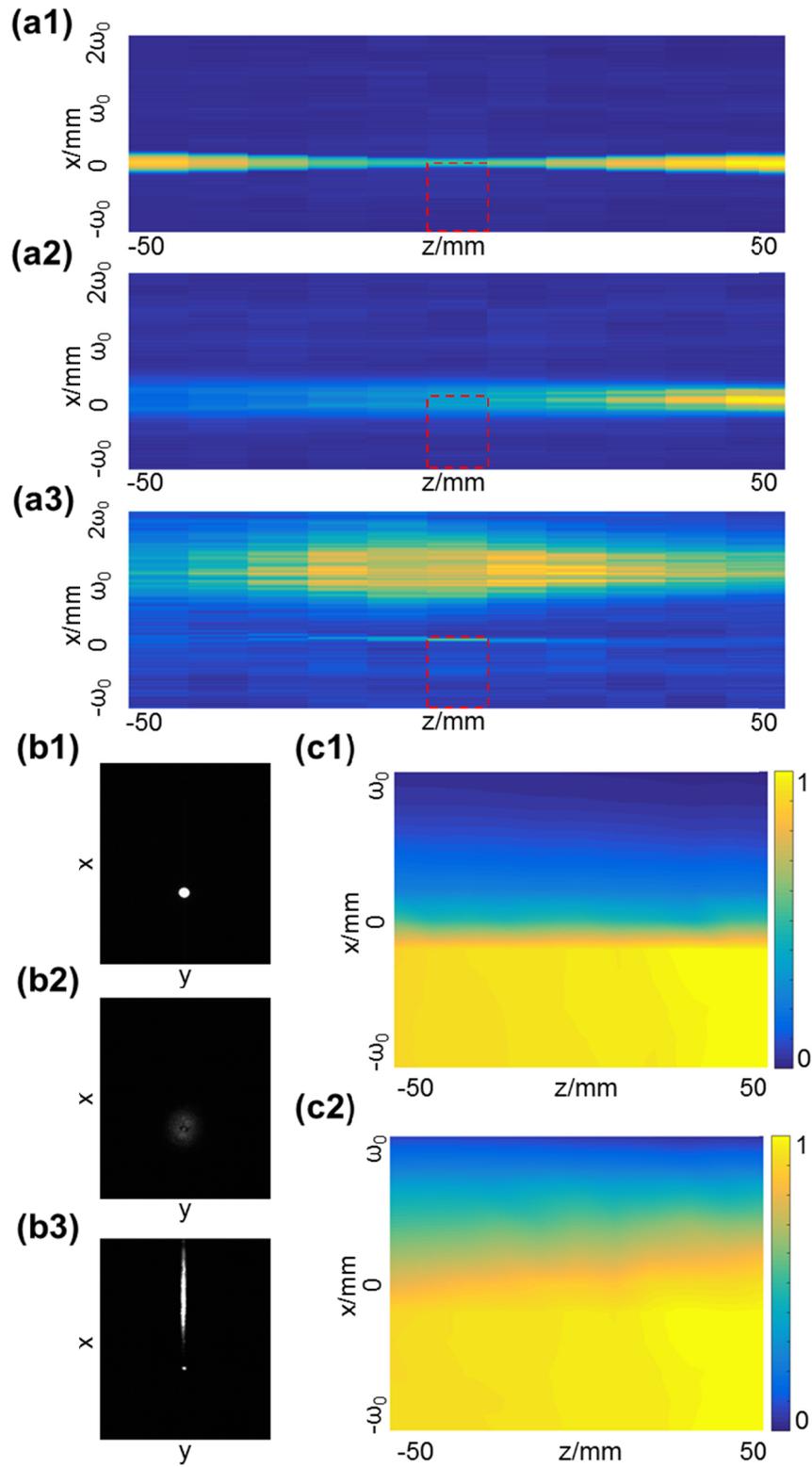

**Fig. 3.** The comparison between trajectories of (a1) original Gaussian beam, (a2) deformed Gaussian beam and (a3) deformed Airy beam. (b1) to (b3) The corresponding beam profiles on the focal plane. The relationship between optimal receiving power and the location of the obstacle top end with (c1) deformed Gaussian beam and (c2) deformed Airy beam.

In each plane, the obstacle is swept in nine equally spaced locations range from $-\omega_0$ to $\omega_0$. The results of the communications performance are shown in Fig. 4. The dashed red horizontal line marks the reference EVR value to achieve bit-error-rate (BER) of 0.0038, which is the soft decision forward error correction (FEC) threshold [25]. The EVM performances of Gaussian and Airy beams have no obvious difference until the block size is more than half of the plane. For both obstacle positions (on or after the focal plane), as the obstacle blocks more of the transverse plane, the Airy beam achieves better performance over the Gaussian beams. For each kind of beam, the Airy beams can also overcome larger obstacles when the obstacle is not on the focal plane, while the Gaussian beams are able to overcome larger obstacles on the focal plane. At the FEC threshold limit, the Gaussian beam can overcome obstacle size of ~ $\omega_0/2$, while the Airy beam is able to evade obstacle size of ~ $\omega_0$. The difference in performance becomes larger when the obstacle is further away from the focal plane.

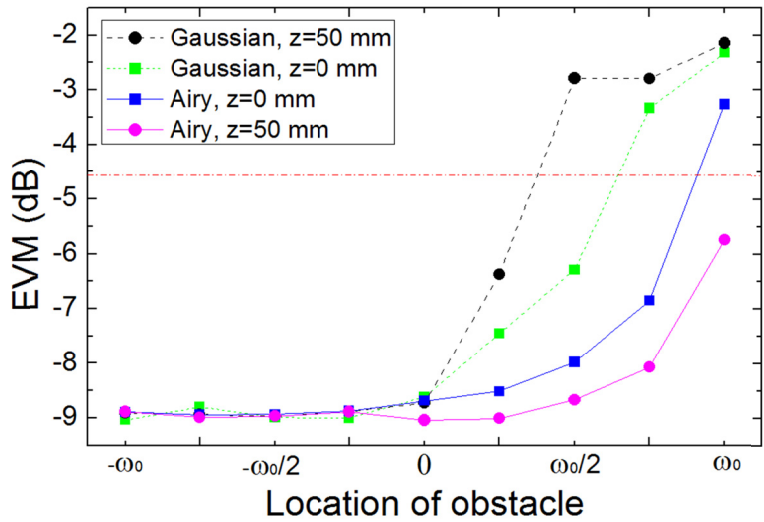

**Fig. 4.** The EVM of the communication system against location of the obstacle with Gaussian and Airy beams.

In this work, we have proposed and demonstrated a scheme of utilizing Airy beams with tailorable ballistic trajectories to overcome obstacles in free space optical communication systems. Two approaches for optimizing the Airy beam trajectory are implemented to evade the preset obstacle represented by an opaque thin plate. Both numerical and experimental comparisons show the obstacle-evading ability of Airy beam is close to the initial beam waist radius and much larger than normal Gaussian beams.

**Funding.** National Natural Science Foundation of China (NSFC) (11774437, U1701661, 61490715, 11690031); National Basic Research Program of China (973 Program) (2014CB340000) Science and Technology Program of Guangzhou (2018-1002-SF-0094); Fundamental Research Funds for the Central Universities of China (SYSU: 17lgzd06).